\documentclass{aa}
\usepackage{txfonts}
\usepackage{graphicx} % when using Latex and dvips 
\usepackage{natbib}
\bibpunct{(}{)}{;}{a}{}{,} % to follow the A&A style
\newcommand{\gsim}{\,\raisebox{0.2em}{$>$}\!\!\!\!\!
\raisebox{-0.25em}{$\sim$}\,}
\newcommand{\lsim}{\,\raisebox{0.2em}{$<$}\!\!\!\!\!
\raisebox{-0.25em}{$\sim$}\,}
\newcommand{\gr}{$\gamma$-ray \,}
\newcommand{\grs}{$\gamma$-rays \,}
\newcommand{\rxj}{RX~J1713.7-3946 }

\begin{document}

\title {Theory of cosmic ray production in the supernova
remnant RX~J1713.7-3946}

   \subtitle{}

   \author{ E.G. Berezhko
          \inst{1}
          \and
          H.J.V\"olk 
          \inst{2}
          }

   \offprints{H.J.V\"olk}

   \institute{Yu.G. Shafer Institute of Cosmophysical Research and Aeronomy,
                     31 Lenin Ave., 677980 Yakutsk, Russia\\
              \email{berezhko@ikfia.ysn.ru}
         \and
           Max Planck Institut f\"ur Kernphysik,
                Postfach 103980, D-69029 Heidelberg, Germany\\
              \email{Heinrich.Voelk@mpi-hd.mpg.de}
             }

   \date{Received month day, year; accepted month day, year}

     \abstract 
     {} 
     {A nonlinear kinetic theory of cosmic ray (CR) acceleration
     in supernova remnants (SNRs) is employed to investigate the properties of
     SNR RX~J1713.7-3946.}  
     {Observations of the nonthermal radio and X-ray
     emission spectra as well as the H.E.S.S. measurements of the very high
     energy \gr emission are used to constrain the astronomical and the
     particle acceleration parameters of the system.}  
     {Under the assumptions
     that \rxj was a core collapse supernova (SN) of type II/Ib with a massive
     progenitor, has an age of $\approx 1600$~yr and is at a distance of
     $\approx 1$~kpc, the theory gives indeed a consistent description for all
     the existing observational data. Specifically it is shown that an
     efficient production of nuclear CRs, leading to strong shock modification,
     and a large downstream magnetic field strength
     $B_{\mathrm{d}}\sim100$~$\mu$G can reproduce in detail the observed
     synchrotron emission from radio to X-ray frequencies together with the \gr
     spectral characteristics as observed by the
     H.E.S.S. telescopes. Small-scale filamentary structures observed in
     nonthermal X-rays provide empirical confirmation for the field
     amplification scenario which leads to a strong depression of the inverse
     Compton and Bremsstrahlung fluxes. Going beyond that and using a
     semi-empirical relation for young SNRs between the resulting CR pressure
     and the amplified magnetic field energy upstream of the outer SN shock as
     well as a moderate upper bound for the mechanical explosion energy, it is
     possible to also demonstrate the actual need for a considerable shock
     modification in \rxj. It is consistent with \rxj being an efficient source
     of nuclear cosmic rays.} 
     {}

   \keywords{cosmic rays -- acceleration of particles
-- supernovae: general -- ISM: individual objects: SNR RX~J1713.7-3946  -- 
radiation mechanisms: non-thermal --
gamma-rays: theory} 
\titlerunning{Theory of CR production in SNR RX~J1713.7-3946}
   \maketitle
%
%________________________________________________________________

\section{Introduction}

RX~J1713-3946 is a shell-type supernova remnant (SNR), located in the Galactic
plane, that was discovered in X-rays with ROSAT \citep{pfef96}.
Further study of this SNR with the ASCA satellite by
\citet{koyama97} and later by \citet{slane99} have shown that the
observable X-ray emission is entirely non-thermal, and this property was
confirmed in later XMM observations \citep{cassam04}.

Given these characteristics it is somewhat surprising that the radio
emission is weak. In fact, only part of the shell could be 
detected in
radio synchrotron emission up to now, with a poorly known spectral form
\citep{laz04}.

RX~J1713-3946 was also detected in very high energy \grs with the
CANGAROO \citep{muraishi00,enomoto02} and H.E.S.S.
\citep{aha04,aha05} telescopes. Especially the latter observations
show a clear shell structure at TeV energies which correlates well with the
ASCA contours.

This intriguing situation has prompted us to theoretically investigate the
acceleration of both electrons and protons in detail, using a nonlinear
kinetic theory. It couples the particle acceleration process
with the hydrodynamics of the thermal gas \citep{byk96,bv00a}.
However, in comparison with the other SNRs
(SN 1006, Cas~A, and Tycho's SNR) that were successfully described within the
framework of this theory and finally allowed a prediction of the \gr flux
\citep{bkv02,vbkr,bkv03,bpv03,bv04a,vbk05,ksen05}, the
case of \rxj puts considerably more obstacles in theory's way. First of all,
such decisive astronomical parameters as source distance, present expansion
velocity and age are not well known. We shall use the values $d=1$~kpc and
$t=1612$~yr (see below). In addition, the lack of knowledge of the spectral
shape of the radio emission makes it difficult to derive -- from synchrotron
spectral observations -- two determining physical quantities: the effective
strength of the magnetic field and the proton injection rate into the diffusive
acceleration process. This technical difficulty adds to the uncertainty about
the type of progenitor star and its possible influence on the circumstellar
medium. At first sight it appears therefore impossible to theoretically predict
the TeV \gr emission. We shall argue nevertheless that the observed overall
synchrotron spectral shape, from radio frequencies to the X-ray cutoff, and the
small-scale filamentary structures in the nonthermal X-ray emission of \rxj are
consistent with efficient CR acceleration associated with a considerably
amplified magnetic field. These properties allow in addition a consistent fit
of the observed TeV \gr spectrum. It is strongly dominated by $\pi^0$-decay
emission.

Assuming that all young SNRs obey a semi-empirical linear correlation between
the amplified magnetic field energy density and the pressure of accelerated
CRs, and that the mechanical explosion energy of \rxj has an upper bound of
$2.5 \times 10^{51}$~erg, also makes it possible to approximately determine the
proton injection rate and thus to predict the time evolution of the CR
energy content of the SNR and the nonlinear backreaction of the CRs on their
own acceleration.

The existence of a compact central object, the X-ray point source 1WGA
J1713.4-3949 -- most probably a neutron star -- and the properties of the
observed nonthermal emission of \rxj strongly indicate that the source is a
core collapse supernova (SN) of type II/Ib which exploded into the 
wind-blown
bubble from a massive progenitor star. This was also argued by \citet{cassam04}.
The main fraction of the remnant volume contains diluted hot bubble
gas, whereas at the current evolutionary phase the outer SN shock already
propagates into the dense shell of the swept-up ambient interstellar medium
(ISM) gas.

The remaining, mainly quantitative uncertainties in the detailed properties of
the circumstellar environment still do not allow a unique theoretical
solution. Therefore our primary aim is to construct a theoretical model that is
consistent in itself and with all available measurements. And this is
indeed possible. Nevertheless, we show that also the prediction of a
highly efficient nuclear CR production can be cogently argued. In short, we
demonstrate that the existing data are consistent with very efficient
acceleration of CR nuclei at the SN shock wave which converts a significant
fraction of the initial SNR energy content into CR energy. We demonstrate in
addition that the available morphological evidence in hard X-rays implies that
the leptonic channels of \gr emission (Inverse Compton emission as well as
Nonthermal Bremsstrahlung radiation) should be strongly suppressed, and that
the calculated hadronic \gr emission is consistent with the H.E.S.S.
spectrum at TeV energies and with the EGRET upper limits in the GeV
range. The further going prediction of efficient acceleration of CR nuclei
on the basis of two additional semi-empirical assumptions is left to the
Discussion section.

\section{Physical parameters of RX~J1713-3946}

Following the estimates of \citet{koyama97} from ASCA X-ray 
observations, and of \citet{fukui03} from NANTEN CO line 
measurements, we adopt a distance $d=1$~kpc. For the present angular 
sources size of 60' this leads to a SNR radius $R_{\mathrm{s}}\approx 
10$~pc. Even though a considerably larger distance 
$d=6$~kpc to the SNR \rxj, estimated in earlier studies \citep{slane99},
is still being discussed \citep[e.g.][]{cassam04,hiraga05}, the recent study of molecular clouds toward SNR \rxj \citep{morig05} ``strongly supports'' the distance of 1~kpc.

A rather simple theoretical consideration shows
that the SNR should be quite young in order to reach such a large size and to
be at the same time such a bright source of nonthermal X-rays. 
The observed energy flux of nonthermal X-rays at 
energy $\epsilon_{\nu}=4$~keV
is $\nu S_{\nu}\approx 100$~eV/(cm$^2$s)
\citep{slane99} which for the distance $d=1$~kpc translates into the
synchrotron luminosity $\nu L_{\nu}=4\pi d^2\nu S_{\nu}\approx
2\times10^{34}$~erg/s. Here $\nu$ is the emission frequency, $L_{\nu}$ and
$S_{\nu}$ are the spectral luminosity and flux, measured at the distance $d$, 
respectively. According to the theoretical
estimates of \citet{bv04b} such a high luminosity 
is possible only if the shock speed is not very low,
$V_{\mathrm{s}}>1500$~km/s. Otherwise the electron cutoff energy becomes 
so small that it
implies a very low X-ray luminosity. This leads to a
restriction on the SNR age for
an assumed uniform external ISM. In such a case the present epoch of SNR
expansion is described by the Sedov solution:
\begin{equation}
R_{\mathrm{s}}=(1.2\times 
10^{24}E_{\mathrm{sn}}/N_{\mathrm{ISM}})^{1/5}t^{2/5},
\label{eq1}
\end{equation}
where $E_{\mathrm{sn}}$ is the SN explosion energy, 
$N_{\mathrm{ISM}}=\rho_{\mathrm{ISM}}/m_{\mathrm{p}}$ is the
interstellar medium (ISM) number density, $\rho_{\mathrm{ISM}}$ is ISM 
density and $m_{\mathrm{p}}$
is the proton mass. Since the shock radius and the shock speed are related as
$V_{\mathrm{s}}=0.4R_{\mathrm{s}}/t$ the above condition gives a 
restriction for the SNR age
$t<2.7\times 10^3$~yr. If we assume the standard value 
$E_{\mathrm{sn}}=10^{51}$~erg for
the SN explosion energy and use the observed SNR radius 
$R_{\mathrm{s}}=10$~pc, Eq.\ref{eq1}
will give us $N_{\mathrm{ISM}}<0.3$~cm$^{-3}$. In such a diluted ISM the 
maximum
possible TeV \gr flux $\epsilon F_{\gamma}(\epsilon)= 10 (N_{\mathrm{ISM}}/1~
\mathrm{cm}^{-3})(d/1~\mathrm{kpc})^{-2}$ eV cm$^{-2}$ s$^{-1}$ 
\citep{bv00b} and the expected nonthermal X-ray flux $\nu
L_{\nu}=10^{33} (K_{\mathrm{ep}}/10^{-4})$~erg/s \citep{bv04b}
would be significantly below the observed values.
Here $K_{\mathrm{ep}}$ denotes the electron:proton ratio in the 
energetic particle population. 

This contradiction can be resolved if SNR \rxj is the result of a core 
collapse SN of type II/Ib which exploded into the adiabatic, very diluted 
bubble created by the wind of a massive progenitor star. This is also 
required by the probable existence of a compact remnant in the center, the 
neutron star 1WGA J1713.4-3949 \citep[e.g.][]{cassam04}.
According to stellar wind theory \citep[e.g.][]{weaver,chev82},
the bubble is bounded by a dense, massive shell which consists of 
the swept-up, cooled ISM material. Note that \citet{slane99}, \citet{fukui03} and \citet{cassam04} came to the same conclusion 
based on different types of arguments \citep[see also][]{ell01}. In a
very general study the combined SNR dynamics and diffusive shock
acceleration in wind-SNe has been investigated by \citet{bv00b}.

Progenitor stars of core collapse SNe, which have intense winds, are massive
main-sequence stars with initial masses $M_\mathrm{i}>15M_{\odot}$ 
\citep[e.g.][]{abb}. In the mean, during their evolution in the
surrounding uniform ISM of gas number density 
$\rho_0=m_\mathrm{p}N_{\mathrm{ISM}}$, they create
a bubble of size \citep{weaver,chev89l}
\begin{equation}
R_{\mathrm{sh}}=0.76( 0.5 \dot{M}V_{\mathrm{w}}^2 
t_{\mathrm{w}}^3/\rho_0)^{1/5},
\label{eq2}
\end{equation}
where $\dot{M}$ is the mass-loss rate of the progenitor, $V_{\mathrm{w}}$ 
is the wind speed, and $t_{\mathrm{w}}$ is the duration of the wind 
period. This bubble could be adiabatic, and then have very low gas 
density, or it could be modified by mixing of the hot bubble gas with 
surrounding shell material. The large size of \rxj and its low age require 
an adiabatic bubble, which we shall adopt for the sequel.

In order to determine the SN shock dynamics inside the shell we model the 
gas number density distribution $N_{\mathrm{g}}$ as a constant in the 
bubble and as a power law in radius in the shell. This leads to the 
following radial profile 
\begin{equation} 
N_{\mathrm{g}}=N_{\mathrm{b}}+(r/R_{\mathrm{sh}})^{3(\sigma_{\mathrm{sh}} 
-1)}N_{\mathrm{sh}}, 
\label{eq3} 
\end{equation} 
where 
$N_{\mathrm{sh}}=\sigma_{\mathrm{sh}} N_{\mathrm{ISM}}$ is the peak number 
density in the shell, $N_{\mathrm{b}}$ is the gas number density inside 
the bubble, typically very small compared with the shell density, and 
$\sigma_{\mathrm{sh}} =N_{\mathrm{sh}}/N_{\mathrm{ISM}}$ is the shell 
compression ratio. We emphasize that the compression ratio 
$\sigma_{\mathrm{\mathrm{sh}}}$ can exceed the adiabatic upper limit of 4 
as a result of the radiative cooling in the shell.

The mass of the bubble 
\begin{equation}
M_{\mathrm{b}}=(4\pi R_{\mathrm{sh}}^3/3)m_{\mathrm{p}}N_{\mathrm{b}}
\label{eq4}
\end{equation}
is rather small, $M_{\mathrm{b}}< M_{\odot}$, in the case of moderate 
progenitor masses
$M_i<20M_{\odot}$ \citep[e.g.][]{chev89l}, whereas the shell mass
\begin{equation}
M_{\mathrm{sh}}=4\pi 
N_{\mathrm{sh}}m_{\mathrm{p}}\int_0^{R_{\mathrm{sh}}}dr 
r^2(r/R_{\mathrm{sh}})^{3(\sigma -1)}=
(4\pi R_{\mathrm{sh}}^3/3)N_{\mathrm{ISM}}m_{\mathrm{p}}
\label{eq5}
\end{equation}
amounts to a few thousand solar masses (see below).

During SNR shock propagation through the adiabatic bubble, only a small
fraction of the mechanical explosion energy is therefore given to gas of
stellar origin. The main fraction of the explosion energy is deposited in the
shell.

Here we use the gas number density distribution 
$N_{\mathrm{g}}(r)=\rho(r)/m_{\mathrm{p}}$
in the form
\begin{equation}
N_{\mathrm{g}}=0.008+1.1[r/(10~\mbox{pc})]^{12}~~\mbox{cm}^{-3}
\label{eq6}
\end{equation}
which gives a consistent fit for all existing data for SNR \rxj. Such a
distribution corresponds to a bubble with $\sigma_{\mathrm{sh}}=5$, 
$N_{\mathrm{b}} =
0.008$~cm$^{-3}$, and a range $17.6 < R_{\mathrm{sh}}< 19.9$~pc, created 
by the
wind of a main-sequence star of initial mass $15M_{\odot}<M_i<20M_{\odot}$ in a
surrounding ISM of number density $110<N_{\mathrm{ISM}}<500$~cm$^{-3}$, 
respectively
\citep{chev89l}. Such a high ISM density, required to produce a
bubble of such a small size, is quite consistent with the value
$N_{\mathrm{ISM}}=300$~cm$^{-3}$ derived from the analysis of
the X-ray emission of SNR \rxj \citep{cassam04}. 
This implies that the bubble is formed inside a molecular cloud; a
discussion of the CO emission from the neighbourhood of the SNR is given in the
paper by \citet{fukui03}.

For the SNR age we use the value $t=1612$~yr, which is consistent with the
hypothesis of \citet{wang97}, based on historical Chinese records, that
RX~J1713-3946 is the remnant of the AD393 ``guest star''. 

We adopt the following values of SNR parameters: 
$E_{\mathrm{sn}}=1.8\times 10^{51}$~erg, $M_{\mathrm{ej}}=3.5M_{\odot}$. 
As shown below, these parameter values lead to a good fit for the observed 
SNR properties. For a better determination of the explosion energy we 
would need the value of the shock speed $V_{\mathrm{s}}$ which is 
unfortunately is not measured.

\subsection{Magnetic field strength}

Whereas the above physical parameters -- explosion type, explosion energy,
progenitor mass loss, circumstellar density, age and distance -- are part of an
internally consistent picture whose detailed quantitative specification allows
a consistent fit to the observed X-ray synchrotron and \gr emission, the
magnetic field strength plays a specific role in this discussion. The reason is
that its strength decides the relative role of electrons and nuclear particles
in the \gr emission. In the following 
we shall therefore go into the question of the field
determination in some detail.

As indicated in the Introduction, there are two obvious ways to determine the
effective magnetic field in a SNR, and we have used both of them in the past
\citep{vbk05,ber05} for SN 1006, Cas~A, and Tycho's SNR.

The first method analyses the overall synchrotron spectrum. If, first of 
all, the radio continuum spectrum is softer than the test particle limit, 
i.e. with a photon index larger than $\alpha=0.5$, then this is attributed 
to the radiation of electrons that have been accelerated at the subshock 
of an overall nonlinearly modified shock. This requires (i) an effective magnetic 
field strength which is above a certain limit and (ii) that the 
electrons have energies below 1~GeV, where their spectrum is steeper than 
$N_{\mathrm{e}}\propto p^{-2}$, in order to obtain such a steep 
synchrotron spectrum \citep[see][for details]{bpv03,ber05}.
In addition, the form of the X-ray synchrotron spectrum, 
including its cutoff, must be consistent with this magnetic field 
strength. This determines the effective field strength inside the SNR to 
within about 30 percent. In all cases considered 
in the past the field turned out 
to be amplified by a factor between 5 and 10 relative 
to the expected upstream field strength \citep{vbk05}.

The second method is also empirical and makes use of the synchrotron 
morphology of the SNR at keV energies \citep{bv04a,vbk05}.
Interpreting the thickness of sharp filaments in 
hard X-rays -- typically in the line-free energy region between 4 and 6 
keV -- as the result of synchrotron losses during and after the electron 
acceleration process, we obtain an independent measure of the downstream 
magnetic field strength. Within an accuracy of 20 to 30 percent this field 
strength agrees with that determined by the first method.

The present status of the radio synchrotron measurements for SNR \rxj does not
determine the radio spectral index, and therefore we have only limited
restrictions on the field strength. Fortunately, however, the synchrotron
morphology gives a rather definite measure of the field. The recent XMM
observations of \citet{hiraga05} show a radial X-ray profile (see their Figs.1
and 2) which we interpret to show unequivocally the filamentary structure
behind the outer SNR shock. The exponential 
angular width of the filament corresponds
to 5.1 bins, of width $\Delta \psi =25.6''$ each (Hiraga, private communication), 
within the
radial profile shown by \citet{hiraga05}. Therefore the profile width
corresponds to $\Delta \psi \approx 2.2'$, 
or to a spatial scale $L=2 \times 10^{18}$~cm at
a distance of 1 kpc.

As it was already demonstrated for other young SNRs \citep{bkv03,bv04a,vbk05},
the measured width of the
projected radial profile of the nonthermal X-ray emission determines 
the internal (downstream) magnetic field according to the expression
\begin{equation}
B_\mathrm{d}=[3m_{\mathrm{e}}^2c^4/(4er_0^2l_2^2)]^{1/3}(\sqrt{1+\delta^2}-
\delta)^{-2/3},
\label{eq7}
\end{equation}
where $\delta^2=0.12[c/(r_0\nu)][V_{\mathrm{s}}/(\sigma c)]^2$, 
$l_2\approx L/7$ is the radial width of the X-ray emissivity 
$q_{\nu}(\epsilon_{\nu},r)$, $r_0$ is the classical electron radius  
and $\sigma$ is the total shock compression ratio. Substituting into 
this expression the values $L=2 \times 10^{18}$~cm, $\sigma=6.3$ 
(see below), $V_{\mathrm{s}}=1800$~km/s, and $\nu =1.8\times 
10^{17}$~Hz ($\epsilon_{\nu}=0.7$~keV), we obtain $B_{\mathrm{d}}\approx 
65$~$\mu$G.

Given the finite resolution used to analyse these XMM data, which 
leads to a wider profile than the actual profile (see Fig.\ref{f5} below), this 
value for $B_{\mathrm{d}}$ is clearly a lower limit to the true 
amplified field. However, the \citet{hiraga05} XMM profile has 
the great advantage that it certainly corresponds to the outer shock, 
whereas for the much narrower $20''$  Chandra profile measured by 
\citet{uchiyama03}, as analysed in \citet{vbk05}, the doubt 
remained, if that profile was due to the outer shock or rather due to some 
other discontinuity in the SNR plasma. On the other hand, there is at 
present only one such profile available, and we are left with the question 
as to how representative this profile and the derived field value is. From 
general gas dynamics considerations we argue that the probability is 
extremely low that the XMM profile corresponds to a singular point at the 
shock surface. We therefore conclude that a lower limit to the magnetic 
field strength at the outer shock is $65~\mu$G. Extrapolating from the 
$20''$ scale of the Chandra profile, the lower field limit could be 
as large as $230 \mu$G. Examinations of other XMM or Chandra
radial profiles are expected to confirm this conclusion.

Below we shall derive the effective field strength as a result of fitting 
the synchrotron spectrum, in particular its X-ray part. For other SNRs 
this fit and the filament sizes give consistent results for the effective 
field. It will be shown that the spectral fit can be achieved with values 
$B_{\mathrm{d}}=126$~$\mu$G as well as $B_{\mathrm{d}}=250$~$\mu$G. These 
values roughly lie in between the two extremes discussed above. Such 
values of $B_{\mathrm{d}}$ are significantly higher than typical ambient 
magnetic fields, also for the considered bubble wall in the molecular 
cloud (the density at the shock is still a factor $\sim 100$ lower than in 
the cloud which implies a strongly decompressed field upstream of the SNR 
shock compared to the value in the cloud). It must be attributed to field 
amplification at the shock front due to the strong wave production by the 
acceleration of CRs far into the nonlinear regime \citep{belll,bell04}.

\section{Model}

Hydrodynamically, a core collapse supernova (SN) explosion ejects a 
shell of matter with total energy $E_{\mathrm{sn}}\approx 10^{51}$~erg and 
mass $M_{\mathrm{ej}}$ equal to a few solar masses. During an initial 
period the shell material has a broad distribution in velocity $v$. The 
fastest part of this ejecta distribution can be described by a power law 
$dM_{\mathrm{ej}}/dv\propto v^\mathrm{2-k}$ \citep[e.g.][]{Jones81,chev82}.
The interaction of the ejecta with the interstellar medium (ISM) 
creates a strong shock there which heats the thermal gas and accelerates 
particles diffusively to a nonthermal CR component of comparable energy 
density. For core collapse SN explosions a value $k=8$ appears appropriate 
which we adopt in this paper.

As pointed out in the Introduction, our theory for this process is based 
on a fully time-dependent, spherically symmetric solution of the CR 
transport equations, coupled nonlinearly with the gas dynamic equations 
for the thermal gas component. Since all relevant equations, initial and 
boundary conditions for this model have already been described in detail 
in these papers, we do not present them here and only briefly discuss the 
most important aspects below (reviewed also by \citet{voelk03} and \citet{ber05}).

The coupling between the ionised thermal gas (plasma) and the energetic 
particles occurs primarily through magnetic field fluctuations carried by 
the plasma which scatter energetic particles in pitch angle and energize 
them, especially in shock waves. The plasma physics of the field 
fluctuations is not worked out yet in full detail. However the 
accelerating CRs effectively excite magnetic fluctuations upstream of the 
outer SN shock in the form of Alfv\'en waves \citep[e.g.][]{bell78,bo78}
as a result of the streaming instability. In quasilinear 
approximation the wave amplitudes $\delta B$ grow to very high amplitudes 
$\delta B > B$, where $B$ is the average field strength \citep{mck82}.
Since these fluctuations scatter CRs extremely strongly, the 
CR diffusion coefficient is assumed to be as small as the Bohm limit 
$\kappa (p)=\kappa (mc)(p/mc)$, where $\kappa(mc)=mc^3/(3eB)$, $e$ and $m$ 
are the particle charge and mass, $p$ denotes the particle momentum, 
and $c$ is the speed of light.

If $B_{\mathrm{ISM}}$ is the pre-existing field in the surrounding medium, 
then the strong streaming instability would suggest that the instability 
growth is restricted by some nonlinear mechanism to the level $\delta 
B\sim B_{\mathrm{ISM}}$, any further turbulent energy being dissipated 
into the thermal gas \citep{vmck81}. The early attempts
to give a full nonlinear description of the magnetic field evolution 
in a numerical simulation \citep{lucb00,belll}
concluded that a considerable amplification of the ``average'' magnetic 
field in the smooth shock precursor -- produced by the finite CR pressure 
-- should occur. Broadly speaking, it was expected that a non-negligible 
fraction of the shock ram pressure $\rho_\mathrm{0} V_\mathrm{s}^2$ 
($\rho_\mathrm{0}$ is the ambient gas density at the current SN shock 
position) is converted into magnetic field energy. Subsequently, \citet{bell04}
argued that this amplification is the result of a nonresonant 
instability, giving rise to what we shall call the effective upstream 
field $B_0>B_{\mathrm{ISM}}$, on top of which the Alfv\'en waves grow to 
amplitudes $\delta B\sim B_0$. The Bohm limit and the gas heating are then 
to be calculated with $B=B_0$.

In our analyses of the synchrotron spectrum of SN 1006, Tycho's SNR, and Cas~A,
such magnetic field amplifications were indeed found; they can only be produced
as a nonlinear effect by a very efficiently accelerated nuclear CR
component. Its energy density, consistent with all existing data, is so high
that it is able to strongly excite magnetohydrodynamic fluctuations, and thus
to amplify the upstream magnetic field $B_\mathrm{ISM}$ to an effective field
$B_\mathrm{0}>B_\mathrm{ISM}$, at the same time to permit efficient CR
scattering on all scales, reaching the Bohm limit. The same large effective
magnetic field turns out to be required, within the errors, by the comparison
of this self-consistent theory with the morphology of the observed X-ray
synchrotron emission, in particular, its spatial fine structure. We shall
therefore also here allow for the possibility of an amplified field.

The number of suprathermal protons injected into the acceleration process 
is described by a dimensionless injection parameter $\eta \ll 1$ which is 
a fixed fraction of the gas particles entering the shock front. For 
simplicity it is assumed that the injected particles have a velocity four 
times higher than the postshock sound speed. 
We have argued before that ion injection is quite efficient at the 
quasiparallel portions of the shock surface, where it is characterised by 
values $\eta=10^{-4}$~to $10^{-3}$ \citep[see][for details]{vbk03}.
Since this injection is expected to be strongly suppressed at the 
quasiperpendicular part of the shock, one should renormalize the results 
for the nucleonic spectrum, calculated within the spherically symmetric 
model. The lack of symmetry in the actual SNR can be approximately taken 
into account by a renormalization factor $f_{\mathrm{re}}<1$, roughly 
$f_{\mathrm{re}}=0.15$~to $0.25$, which diminishes the nucleonic CR 
production efficiency, calculated in the spherical model, and all effects 
associated with it.

We assume that also electrons are injected into the acceleration process 
at the shock front. Formally their injection momentum is taken to be the 
same as that of the protons. Since the details of the electron injection 
process are poorly known, we chose the electron injection rate such that 
the electron:proton ratio $K_\mathrm{ep}$ (which we define as the ratio of 
their distribution functions at all rigidities where the protons are 
already relativistic and the electrons have not been yet cooled 
radiatively) is a constant to be determined from the synchrotron 
observations. It is demonstrated below, that in the case of SNR \rxj the 
scarce existing data nevertheless give a possibility to estimate the 
values of $B_0$, $\eta$ and $K_{\mathrm{ep}}$.

The electron dynamics is exactly the same as that for protons for electron
rigidities corresponding to ultrarelativistic protons, as long as synchrotron
losses are neglected. Therefore, beyond such rigidities and below the loss
region the distribution function of accelerated electrons has the form
$f_\mathrm{e}(p)= K_\mathrm{ep} f(p)$ at any given time.  The electron
distribution function $f_\mathrm{e}(p)$ deviates only at sufficiently large
momenta from this relation due to synchrotron losses, which are taken into
account by supplementing the ordinary diffusive transport equation by a
radiative loss term.

Clearly, from the point of view of injection/acceleration theory, we must treat
$K_\mathrm{ep}$, together with $B_0$ and $\eta$, as a
theoretically not very well constrained parameter to be quantitatively
determined by comparison with synchrotron observations to the extent that they
are available.

The solution of the dynamic equations at each instant of time yields the CR
spectrum and the spatial distributions of CRs and thermal gas. This allows
the calculation of the expected fluxes of nonthermal emission produced by the
accelerated CRs.

\section{Results} 

\begin{figure}
\centering
\includegraphics[width=7.5cm]{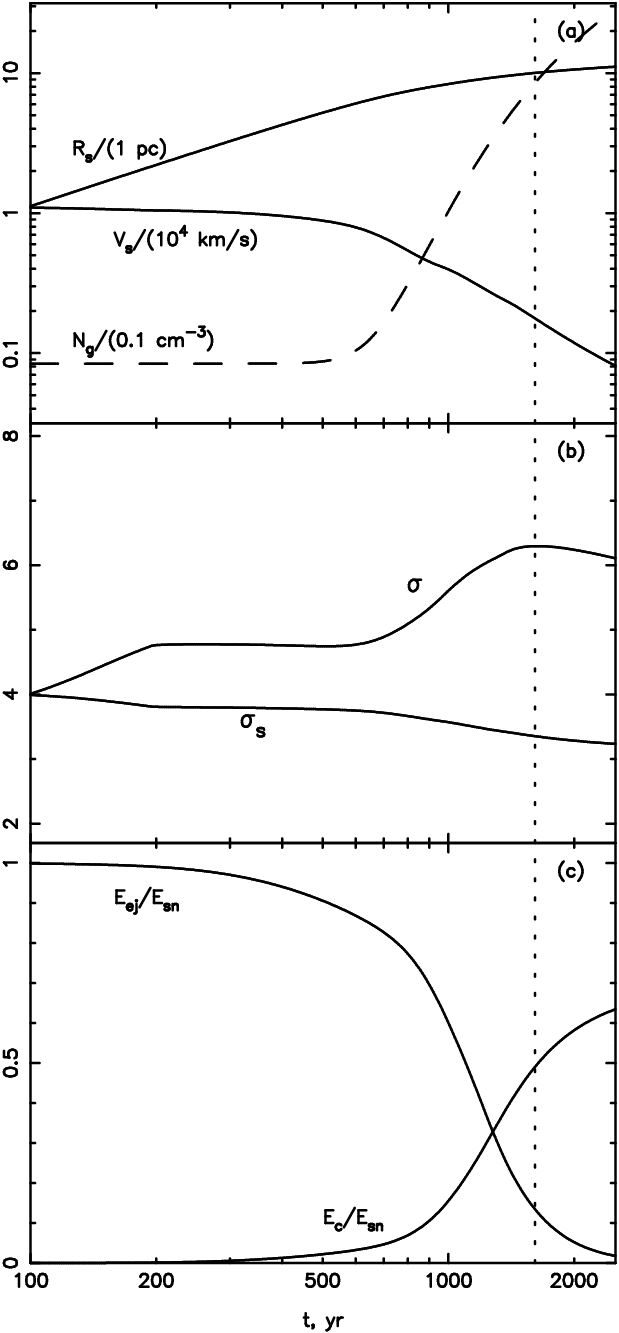}
\caption{(a) Shock radius $R_\mathrm{s}$ and shock speed
  $V_\mathrm{s}$; (b) total shock ($\sigma$) and subshock ($\sigma_\mathrm{s}$)
  compression ratios; (c) ejecta ($E_\mathrm{ej}$) and CR ($E_\mathrm{c}$)
  energies as a function of time, normalised to the total mechanical SNR energy
  $E_\mathrm{sn}$. The vertical dotted line marks the current evolutionary
  epoch.} 
\label{f1}
\end{figure}

The computed gas dynamical characteristics of the SNR are shown in
Fig.\ref{f1}. They demonstrate in Fig.\ref{f1}a that the calculation fits the observed SNR
size $R_{\mathrm{s}}$ for the assumed distance and SNR age. The resulting 
shock
velocity is $V_\mathrm{s} \approx 1840$~km/s.

%------------------------------------------------------------------------fig.1-
%------------------------------------------------------------------------------

To obtain a good fit for the observed synchrotron and gamma-ray spectra 
(see below) we assume a proton injection rate $\eta=3\times 10^{-4}$. We 
want to emphasise that the most reliable way to determine $\eta$ is based 
on the measured spectral shape of the radio emission. Unfortunately this 
can not be done for the case of \rxj, because the shape of the radio 
spectrum is poorly known. The assumed injection rate is nevertheless in 
the same range as in other remnants, where this quantity can indeed be 
directly determined. It leads to a significant nonlinear modification of 
the shock which, at the current age of $t=1612$~yrs, has a total 
compression ratio $\sigma=6.3$ and a subshock compression ratio 
$\sigma_{\mathrm{s}}=3.3$ (Fig.\ref{f1}b). In the present case the shape of 
the TeV-spectrum (see below) is the only spectral characteristic which 
furnishes evidence that the SN shock is strongly modified.

We note here that the lack of a restriction on the injection rate, 
resulting from the lack of knowledge of the radio spectral index,
facilitates the fit of the spectrum. However, we will argue in section 5 
that no basically different injection rate is possible.

For its adopted density the hot wind bubble contains only a small amount 
of gas $M_\mathrm{b}\approx 0.3M_{\odot}$. This is consistent with the 
range of $0.07 < M_\mathrm{b} < 0.6M_{\odot}$ for the initial stellar 
masses $M_\mathrm{i}$ in the range $15M_{\odot} < M_{\mathrm{i}} < 
20M_{\odot} $. Due to this fact the SN shock has deposited less than 20\% 
of the explosion energy while propagating through the bubble ($t<700$~yr), 
as seen from Fig.\ref{f1}c. Yet at the current epoch the SN shock has already 
swept up a considerable mass $M_{\mathrm{sw}}\approx 20M_{\odot}$ and 
therefore the ejecta has already transformed about 90\% of its initial 
energy into thermal gas and CRs.

In spherical symmetry the acceleration process is then characterised by a high
efficiency: at the current time about 53\% of the explosion energy have been
transferred to CRs, and the CR energy content $E_\mathrm{c}$ continues to 
increase to a
maximum of about 60\% in the later phase (Fig.\ref{f1}c), when the highest energy
particles start to leave the source. As typically predicted by the spherically
symmetric model, such a CR acceleration efficiency is significantly higher than
required for the average replenishment of the Galactic CRs by SNRs,
corresponding to $E_{\mathrm{c}}\approx 0.1E_{\mathrm{sn}}$. And as 
discussed above, the necessary
deviation of the SNR from spherical symmetry requires a renormalization of the
number of hadronic CRs. We choose the value $f_{\mathrm{re}}=1/5$.

With this renormalization the CRs inside RX~J1713-3946 SNR already contain
\begin{equation}
E_{\mathrm{c}}=0.53f_{\mathrm{re}}E_{\mathrm{sn}}\approx2\times10^{50}~\mbox{erg}.
\label{eq8}
\end{equation}
% 
%------------------------------------------------------------------------fig.2-
\begin{figure} 
\centering 
\includegraphics[width=7.5cm]{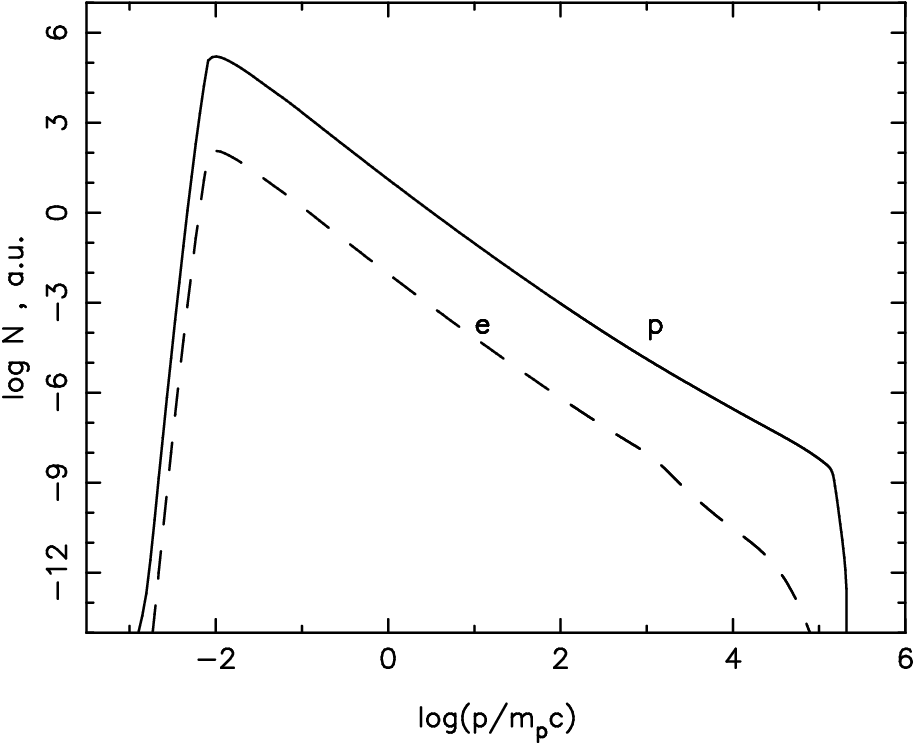}
\caption{Spatially integrated CR spectrum as function of particle momentum. 
  Solid and dashed lines correspond to protons and electrons,
  respectively.}
\label{f2} 
\end{figure}
%------------------------------------------------------------------------------

The volume-integrated (or overall) CR spectrum
 \begin{equation}
N(p,t)=16\pi^2p^2 \int_0^{\infty}dr r^2 f(r,p,t)  
\label{eq9}
\end{equation} 
has, for the case of protons, almost a pure power-law form $N\propto
p^{-\gamma}$ over a wide momentum range from $10^{-2}m_\mathrm{p}c$ up to the
cutoff momentum $p_\mathrm{max}\approx 2\times 10^5m_\mathrm{p}c$ 
(Fig.\ref{f2}). This
value $p_\mathrm{max}$ is limited mainly by 
the finite size and speed of the shock, its deceleration and the
adiabatic cooling effect in the downstream region \citep{ber96}. As a
result of the shock modification, the power-law index slowly varies from
$\gamma=2.2$ at $p\,\,\raisebox{0.2em}{$<$}\!\!\!\!\!
\raisebox{-0.25em}{$\sim$}\,\, m_\mathrm{p}c$ to $\gamma=1.7$ at
$p\,\,\raisebox{0.2em}{$>$}\!\!\!\!\!
\raisebox{-0.25em}{$\sim$}\,\,10^3m_\mathrm{p}c$.

The shape of the overall electron spectrum $N_\mathrm{e}(p)$ deviates from that
of the proton spectrum $N(p)$ at high momenta $p>p_\mathrm{l}\sim
10^3m_\mathrm{p}c$, on account of the synchrotron losses during their residence
time in the downstream region with a magnetic field strength
$B_\mathrm{d}\approx 130~\mu$G which is assumed uniform in this region
($B_\mathrm{d}=B_2=\sigma B_0$). Cf. \citet{bkv02} the synchrotron
losses become important for electron momenta greater than
\begin{equation} 
\frac{p_\mathrm{l}}{m_\mathrm{p}c} \approx 
1.3 \left(\frac{10^8~\mbox{yr}}{t}\right)
\left(\frac{10~\mu\mbox{G}}{B_\mathrm{d}}\right)^2, 
\label{eq10}
\end{equation}
where $p_\mathrm{l}$ is relevant only in the evolutionary stage, when it 
becomes lower than the electron cutoff momentum 
$p_{\mathrm{max}}^{\mathrm{e}}$, $p_\mathrm{l}<p_\mathrm{max}^\mathrm{e}$. 
This is already the case at present. Substituting the SN age $t=1612$~yr 
into this expression, we have $p_\mathrm{l}\approx 500~m_\mathrm{p}c$, in 
agreement with the numerical results (Fig.\ref{f2}).

The shock continuously produces an electron spectrum $f_\mathrm{e}\propto
p^{-q}$, with $q \approx 4$, up to the maximum momentum
$p_\mathrm{max}^\mathrm{e}(t)$ which is at the present time already much
larger than $p_\mathrm{l}$.  Therefore, within the momentum range
$p_\mathrm{l}< p < p_\mathrm{max}^\mathrm{e}$, the electron spectrum is
considerably steeper $N_\mathrm{e}\propto p^{-3}$ due to synchrotron losses
taking place in the downstream region after the acceleration at the shock
front.
%------------------------------------------------------------------------fig.3-
\begin{figure*}
\centering
\includegraphics[width=17.cm]{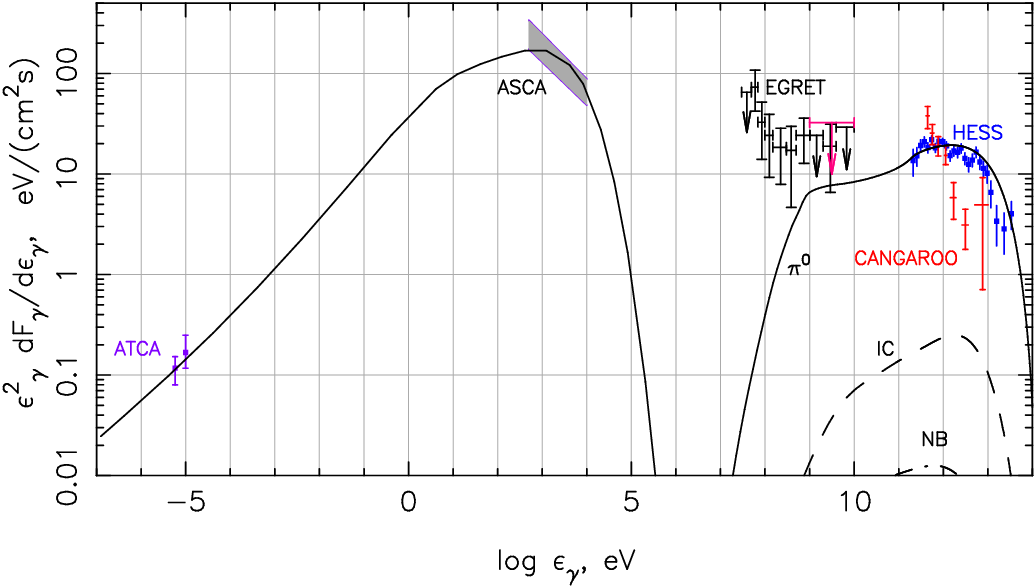}
\caption{Spatially integrated spectral energy distribution of \rxj. The 
{ATCA} radio data \citep[cf.][]{aha05}, {ASCA} X-ray data
\citep[cf.][]{aha05}, {EGRET} spectrum of 3EG~J1714-3857 \citep{reimerp02},
{CANGAROO} data \citep{enomoto02}, in red color) and
{H.E.S.S.} data \citep{aha05}, in blue color) are
shown. The {EGRET} upper limit for the \rxj position \citep{aha05}
is shown as well (red colour). The solid curve at
energies above $10^7$~eV corresponds to $\pi^0$-decay \gr emission, whereas the
dashed and dash-dotted curves indicate the inverse Compton (IC) and
Nonthermal Bremsstrahlung (NB) emissions, respectively.}
\label{f3}
\end{figure*}
%------------------------------------------------------------------------------

For $p_\mathrm{max}^\mathrm{e}< p_\mathrm{max}$ the maximum electron
momentum can be estimated by equating the synchrotron loss time and the
acceleration time. This gives \citep[e.g.][]{bkv02}:
\[
\frac{p_\mathrm{max}^\mathrm{e}}{m_\mathrm{p}c}
= 6.7\times 10^4 \left(\frac{V_\mathrm{s}}{10^3~\mbox{km/s}}\right)
\]
\begin{equation}
\hspace{1cm}\times
\sqrt{\frac{(\sigma-1)}{\sigma (1+ \sigma^2)}
  \left(\frac{10~\mu\mbox{G}}{B_0}\right)}. 
  \label{eq11}
\end{equation}
At the current epoch $V_\mathrm{s}\approx 1840$~km/s which leads to a maximum
electron momentum $p_\mathrm{max}^\mathrm{e}\approx 1.3\times
10^4m_\mathrm{p}c$, in agreement with the numerical results (Fig.\ref{f2}).

We note that during the last thousand years the SN shock speed has been 
going down rapidly. In previous evolutionary epochs it has therefore 
produced electron spectra with cutoff momenta $p_\mathrm{max}^\mathrm{e}$ 
larger than that of the current epoch. Due to this fact the overall 
electron spectrum has a relatively smooth cutoff, extending up to $p\sim 
10^5m_\mathrm{p}c$ (see Fig.\ref{f2}). Together with the synchrotron cooling it 
provides a very good fit of the observed X-ray spectrum (see below).

The parameters $K_\mathrm{ep}\approx 10^{-4}$ and $B_\mathrm{d}= 126$~$\mu$G
give good agreement between the calculated and the measured synchrotron
emission in the radio to X-ray ranges (Fig.\ref{f3}). The steepening of the
electron spectrum at high energies $> 10^3m_\mathrm{p}c^2$ due to synchrotron
losses and the smooth cutoff of the overall electron spectrum naturally yield a
fit to the X-ray data with their soft spectrum. 

The overall broadband spectral energy distribution is displayed in
Fig.\ref{f3}, together with the experimental data from {ATCA} at radio
wavelengths, as estimated for the full remnant by \citet{aha05}, the X-ray data
from {ASCA}, the {EGRET} spectrum of the nearby source 3EG~J1714-3857
\citep{reimerp02}, and the TeV \gr spectra from {CANGAROO} \citep{enomoto02}
and {H.E.S.S.} \citep{aha05}.  The data also include the more recent {EGRET}
upper limit based on the assumption that 3EG~J1714-3857 is not physically
associated with \rxj. The overall fit is impressive, noting that the choice of
a few key parameters like $\eta$, $B_\mathrm{d}$, and $E_{\mathrm{sn}}$ in the
theory allows a spectrum determination over more than 19 decades. The remainder
of this section will give a detailed discussion of these spectra and of the
morphology.

\begin{figure}
\centering
\includegraphics[width=7.5cm]{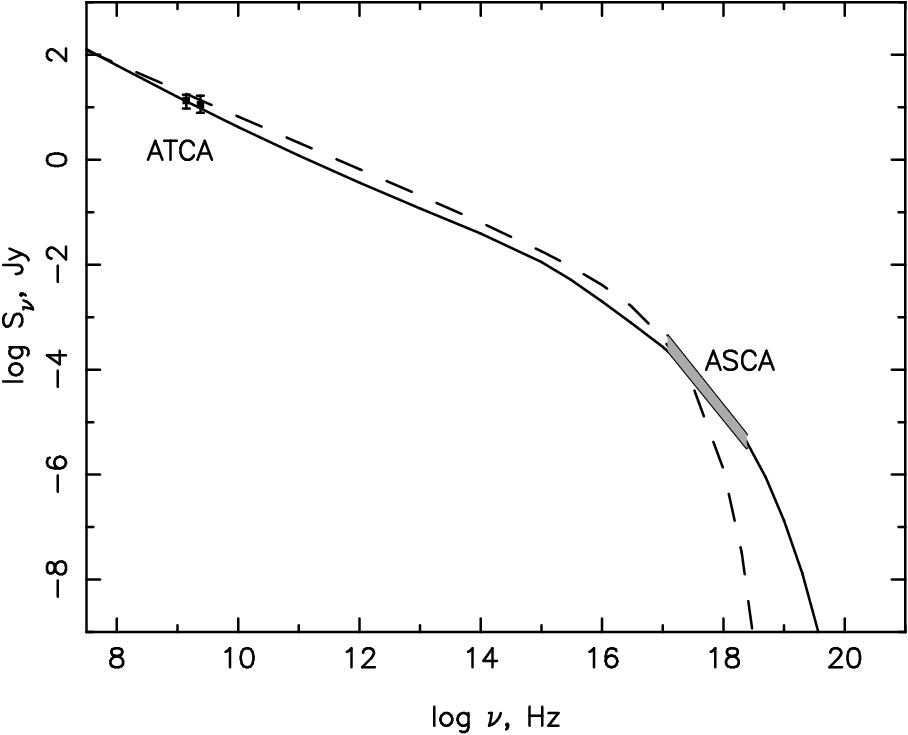}
\caption{Synchrotron photon flux density as a
function of frequency. The solid line corresponds to the 
high-injection model, the dashed line corresponds to
the test-particle approximation. The {ATCA} radio
data and {ASCA} X-ray data \citep{aha05} are shown.}
\label{f4}
\end{figure}

In Fig.\ref{f4} we separately present the synchrotron spectrum, produced at the
current epoch by the accelerated electrons. For comparison we also include
(through the dashed curve) a synchrotron spectrum, which would correspond to an
artificial scenario with a proton injection rate so small ($\eta=10^{-5}$) that
the accelerated nuclear CRs do not produce any significant shock modification
and therefore also no magnetic field amplification. This corresponds to the
test particle limit, and the low value $B_0=5$~$\mu$G was adopted for this
case, for which synchrotron cooling is negligible. There are two meager
differences in the synchrotron spectra, corresponding to these two
scenarios. The high-injection scenario leads to a steep radio spectrum
$S_{\nu}\propto \nu^{-\alpha}$ with power law index $\alpha=0.62$ whereas in
the test particle case $\alpha=0.5$. The quality of the existing radio data
does not allow to distinguish these two scenarios. On the other hand the two
spectra behave essentially different at X-ray frequencies $\nu\gsim
10^{18}$. They demonstrate that only in the high-injection case case with its
high, amplified magnetic field value $B_\mathrm{d}\approx 100$~$\mu$G the 
spectrum
$S_{\nu}(\nu)$ has a smooth cutoff consistent with the experiment (see
Fig.\ref{f4}). In the test particle case the spectrum $S_{\nu}(\nu)$ has too sharp a
cutoff to be consistent with the observations.

The properties of small scale structures of SNR \rxj seen in X-rays by
\citet{uchiyama03}, and in particular by \citet{hiraga05}, provide even 
stronger evidence that the magnetic field inside the SNR is indeed considerably
amplified (see section 2.1).

\begin{figure}
\centering
\includegraphics[width=7.5cm]{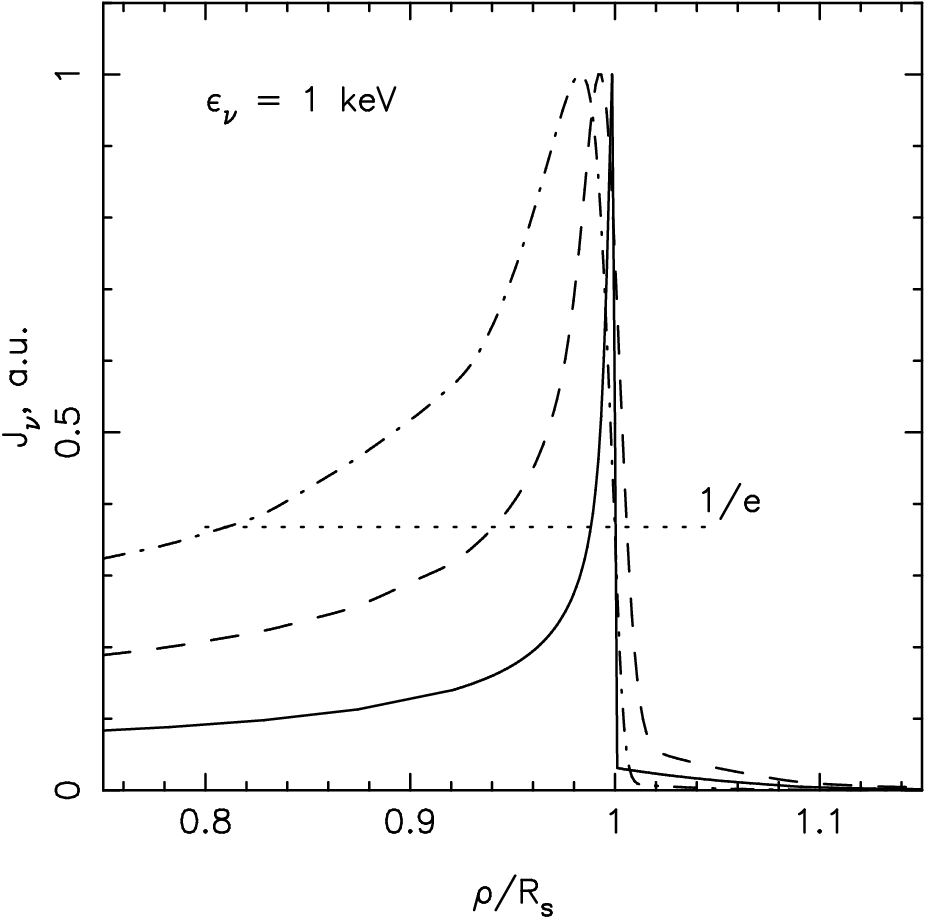}
\caption{Projected radial profile of the X-ray synchrotron emission 
for the energy $\epsilon_\mathrm{\nu}=1$~keV. The solid line corresponds
to the high-injection model; the dashed line again represents the above
profile, but smoothed to the resolution of the {XMM-Newton} data used in
\citet{hiraga05}; the dash-dotted line corresponds to the
test-particle limit. For purposes of presentation all profiles are normalised
to their peak values in this figure.}
\label{f5}
\end{figure}

In order to find out if the filamentary structure found
by \citep{hiraga05} is indeed consistent with the high-injection case we
show in Fig.\ref{f5} the projected radial brightness profile
\begin{equation}
J(\epsilon,\rho)\propto 
\int dx q(\epsilon,r=\sqrt{\rho^2+x^2},x),
\label{eq12}
\end{equation}
calculated for the X-ray energy 
$\epsilon=\epsilon_{\nu}=1$~keV. Here $q(\epsilon, r)$
is the spectral luminosity of the nonthermal emission with photon energy
$\epsilon$. The integration is performed along the line of sight $x$. For high
injection the theory predicts the peak of the emission just behind the shock
front with thickness $\Delta \rho/R_\mathrm{s}\approx 10^{-2}$ that 
corresponds to an
angular width $\Delta \psi\approx 0.4'$. This width is significantly thinner
than the observed width of $\Delta \psi\approx 2.2'$. 
However, one should take into
account the smoothing procedure with which the data were
obtained/presented. Instead of the actual profile $J_{\gamma}(\rho)$ it gives a
broadened profile
\begin{equation}
J'(\rho)=\int_{-\infty}^{\infty}d\rho'
G(\rho,\rho')J(\rho'),
\label{eq13}
\end{equation}
through the (Gaussian) point spread function
\begin{equation}
G(\rho,\rho')=(\sigma_{\rho}\sqrt{2/\pi})\exp[-(\rho-\rho')^2/(2\sigma_{\rho}^2)],
\label{eq14}
\end{equation}
where $\sigma_{\rho}=\sigma_{\psi}d$, and $\sigma_{\psi}$ is the angular
resolution of the instrument (in radians). If we apply such a point spread
function with $\sigma_{\psi}=12.8''$ (corresponding to the smoothed {XMM}
data) to the calculated profile, we obtain the broadened profile , which is
shown in Fig.\ref{f5} by the dashed curve.  It has a width $\Delta 
\rho/R_\mathrm{s}\approx 6.6
\times 10^{-2}$, or $\Delta \psi\approx 2'$, and is consistent with
the observational profile presented in Fig.2 of \citet{hiraga05} which was
discussed in section 2.1.
%------------------------------------------------------------------------fig.4-
%------------------------------------------------------------------------------
%

In addition we present in Fig.\ref{f5} the projected radial profile (dash-dotted 
line) which corresponds to the test particle limit. Since in this case the 
synchrotron losses are not relevant on account of the much lower magnetic 
field value $B_\mathrm{d}=20$~$\mu$G, the peak of the emission is broader 
by a factor of five, inconsistent with the {XMM} observations.

Morphologically the remnant has a rather complex structure. This is
particularly visible in the very bright western parts. Instead of a single,
bright thin rim which should be identified with the actual position of the
forward SNR shock, there are two nested, narrow arc-like rims \citep{cassam04}.
The additional second inner bright rim can not be associated
within our model with a structure situated inside the remnant at the
corresponding distance behind the forward shock. On the other hand
\citet{cassam04} found a positive correlation between the X-ray
brightness and the absorbing column density. It is possibly explained by
assuming that all bright spots observed in X-rays lie just behind the forward
shock, whose shape is significantly distorted due to distortions of the dense,
swept-up shell. Such shell distortions can in turn can be expected because of
inhomogeneities of the external ISM. In such a picture the outer rim,
associated with the position of the forward SNR shock, is bright due to the
projection effect, whereas the inner rim is bright because the corresponding
part of the shock interacts with a denser medium.
%------------------------------------------------------------------fig.5-
%-------------------------------------------------------------------------
%

\begin{figure}
\centering
\includegraphics[width=7.5cm]{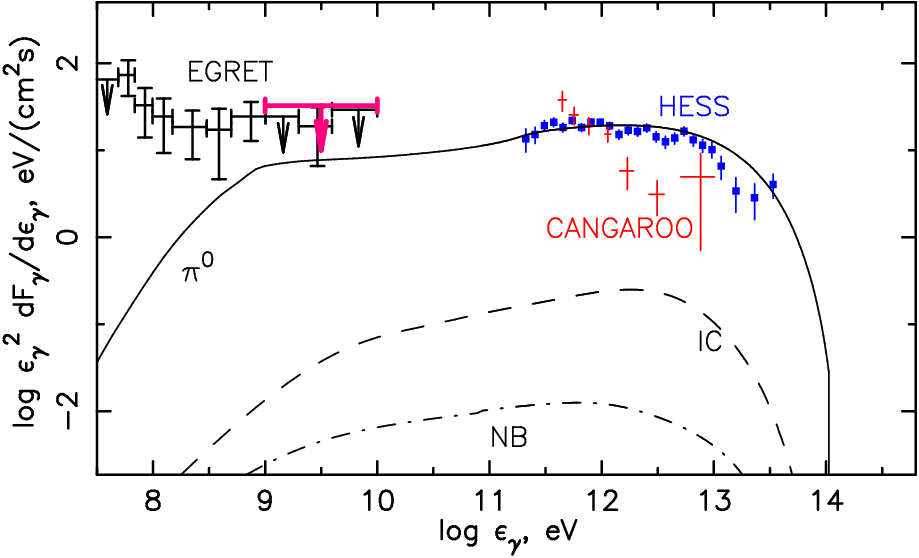}
\caption{Nonthermal bremsstrahlung (dash-dotted line)
IC (dashed line) and $\pi^0$-decay (solid line) $\gamma$-ray
spectral energy distribution as a function of $\gamma$-ray energy.  The
observed H.E.S.S. \citep{aha05} and CANGAROO \citep{enomoto02}
$\gamma$-ray fluxes and the EGRET spectrum of 3EG J1714-3857
\citep{reimerp02} and the EGRET upper limit are shown as well.}
\label{f6}
\end{figure}

Concentrating now on $\gamma$-ray energies the calculated nonthermal 
Bremsstrahlung, IC and $\pi^0$-decay $\gamma$-ray spectral energy 
distributions are shown in Fig.\ref{f6} together with the existing 
experimental data. (In this presentation the theoretical spectrum can 
be directly visually compared with the H.E.S.S. results, cf. Fig.20 of
\citet{aha05}.) According to the calculation, the hadronic 
$\gamma$-ray production exceeds the electron contribution by more than two 
orders of magnitude at all energies (Fig.\ref{f6}).

In the context of the {CANGAROO} observations of \rxj, which covered only
the northern part of the remnant \citep{enomoto02}, \citet{butt02} as
well as \citet{reimerp02} had argued that the GeV emission from the nearby
unidentified {EGRET} source 3EG~J1714-3857 is either associated with the
SNR or an upper limit to the $\gamma$-ray emission of \rxj. As one can see from
Fig.\ref{f6} such an upper limit does not contradict our calculated spectrum of
\rxj. This is even true for an extrapolation of the observed {H.E.S.S.}
spectrum in the test-particle approximation, with a spectral index of 
$\approx 2$ \citep{aha05}.

For energies $\epsilon_{\gamma}<3$~TeV the theoretical \gr spectrum is as 
hard as $dF_{\gamma}/d\epsilon_{\gamma}\propto \epsilon_{\gamma}^{-1.8}$, 
whereas for $\epsilon_{\gamma}>10$~TeV it has a smooth cutoff. We note 
that the \gr cutoff energy $\epsilon_{\gamma}^{\mathrm{max}}\approx 
0.1~cp_{\mathrm{max}}$ is sensitive to the magnetic field strength 
$B_\mathrm{d}$, since the proton cutoff momentum has a dependence 
$p_{\mathrm{max}}\propto R_\mathrm{s}V_\mathrm{s}B_\mathrm{d}$ \citep{ber96}.
It is clearly seen from Fig.\ref{f6} that the calculated spectrum in a 
satisfactory way fits the {H.E.S.S.} measurements.

Since the SN shock propagates through the shell with a rising gas density, the
TeV emission is expected to increase with a rate of about 0.4~\%/yr, whereas
the nonthermal X-ray emission goes down with a rate 0.1~\%/yr due to the shock
deceleration.

We note also that within our approach there is no need to suggest the
interaction of the SN shock with the molecular cloud to be followed by a strong
escape of high energy CRs, in order to explain the observed steepening of
\gr spectrum at $\epsilon_{\gamma}> 1$~TeV, as proposed by \citet{malkov05}.
%------------------------------------------------------------------------fig.6-
%------------------------------------------------------------------------------

\begin{figure}
\centering
\includegraphics[width=7.5cm]{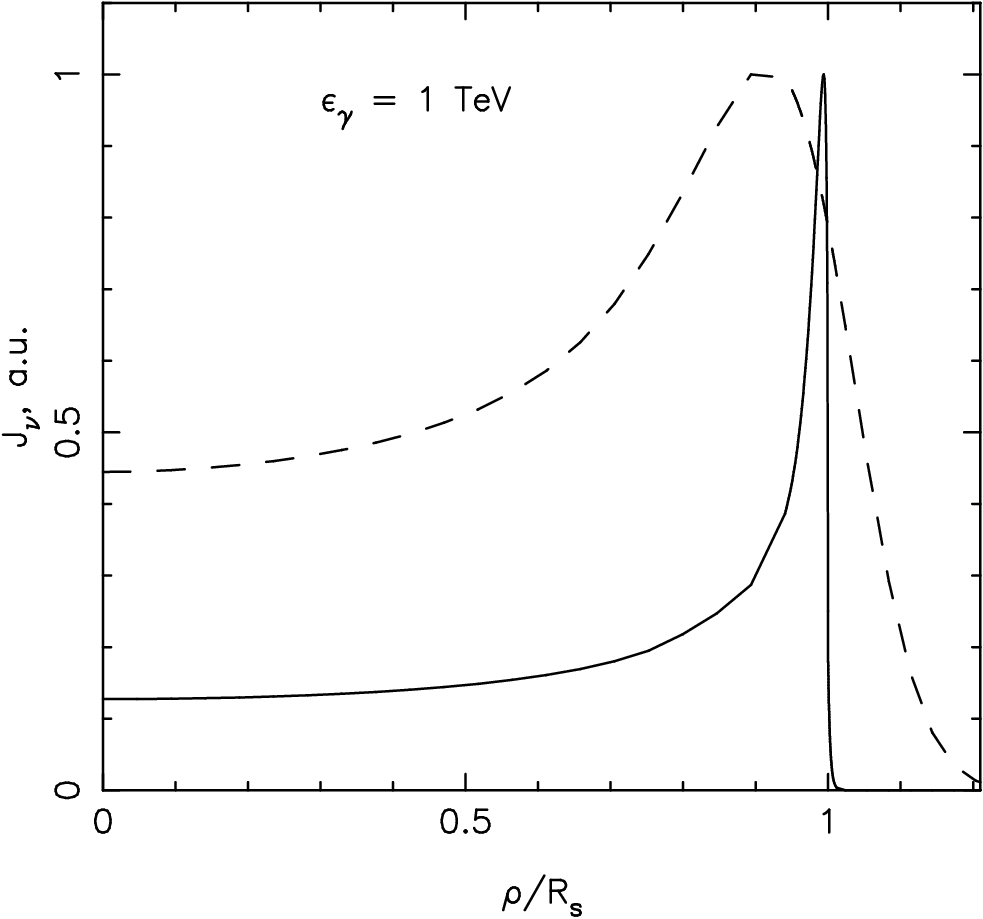}
\caption{The \gr emissivity for \gr energies $\epsilon_\mathrm{\gamma}=1$~TeV
 as a function of angular radial distance. The calculated radial profile is
 given by the solid line; the dashed line represents the calculated
 profile smoothed with a Gaussian point spread function of the width
 $\sigma_{\psi}=0.05^\circ$. For purposes of presentation both profiles are
 normalised to their peak values in this figure.}
\label{f7}
\end{figure}

The projected radial $\gamma$-ray emission profile, calculated for the energy
$\epsilon_\mathrm{\gamma}=1$~TeV, is presented in Fig.\ref{f7}. As a result of the
large radial gradient of the gas density and the CR distribution inside the
SNR, the theoretically predicted radial profile of the TeV-emission is
concentrated within a thin shell of width $\Delta \rho\approx 
0.1R_\mathrm{s}$. Due to
the projection effect this width is seven times larger than the width of the
3-dimensional radial emissivity profile, which is as thin as $\Delta r\approx
0.01R_\mathrm{s}$. Since the H.E.S.S. instrument has a finite 
angular resolution we
present in Fig.\ref{f7} also the modified radial profile smoothed with the Gaussian
point spread function with $\sigma_{\rho}=\Delta \rho= 0.08R_\mathrm{s}$, 
that
corresponds to the angular resolution $2\sigma_{ \psi}=0.1^ \circ$.
%
%------------------------------------------------------------------fig.7-
%-------------------------------------------------------------------------
%
As shown by Fig.\ref{f7} the smoothed radial profile of the TeV-emission is much 
broader and -- what is most interesting -- it is characterised by a 
maximum to minimum intensity ratio 
$J^{\mathrm{max}}_{\gamma}/J^{\mathrm{min}}_{\gamma} =2.2$. Such a ratio 
is consistent with the H.E.S.S. measurement \citep{aha05}
which obviously gives only a lower limit to the sharpness of the 
$\gamma$-ray profile. We conclude that the broad radial profile of the 
TeV-emission with $J^{\mathrm{max}}_{\gamma}/J^{\mathrm{min}}_{\gamma} 
\approx 2$ measured by the H.E.S.S. instrument is indirect evidence 
that the actual radial profile is significantly sharper, with a higher 
ratio $J^{\mathrm{max}}_{\gamma}/J^{\mathrm{min}}_{\gamma} > 2.2$. Also, 
comparing with the 
radial X-ray profile in Fig.\ref{f5}, the similarity of the radial profiles of 
the X-ray, as measured by ASCA and the radial TeV \gr profiles, as 
obtained by H.E.S.S., must be seen as a consequence of the finite 
angular resolutions of these instruments. In reality, the X-ray profiles 
should be narrower than the \gr profiles due to the synchrotron losses.

\section{Discussion and summary}

The theoretical results presented here are fully consistent with a dominantly
hadronic origin of the observed TeV \grs. Beyond that they of course imply that
the overwhelming contribution to the nonthermal pressure comes from accelerated
nuclear particles. Despite the observational fact that the broadband emission
is not fully spherically symmetric, with the brightness increasing from
southeast to northwest, the quasi-spherical character of the X-ray rims on the
western half of the SNR suggests the approximate validity of a spherically
symmetric model, with clear limitations due to the non-spherical nature of the
magnetic field configuration which are to be corrected for. 

The true difficulty for the theoretical description is therefore not the 
complex geometry of the distribution of molecular gas and X-ray absorbing 
column density, but rather the fact that several key parameters of this 
source are either not known or poorly constrained. This already concerns 
the distance and age of the object. We follow present consensus which puts 
the distance at 1 kpc and the age to about 1600 years. We also concur with 
other authors and argue that the primary explosion must have been a type 
II/Ib SN event with a massive progenitor star whose mass loss in the main 
sequence phase created a hot wind bubble in a high-density environment. 
However, at present these remain assumptions, as well-founded as they are. 
The solution for the overall remnant dynamics then yields the value for 
the expansion velocity of the outer shock, given the total mechanical 
energy $E_{\mathrm{sn}}$ released in the explosion. To obtain a consistent 
solution for the broadband nonthermal emission we chose $E_{\mathrm{sn}}= 
1.8 \times 10^{51}$~erg which is already on the high side, also for core 
collapse SNRs with initial main sequence progenitor masses between 15 and 
20$M_{\odot}$ \citep{Hamuy03}. Therefore this particular parameter value is 
limited from above.

The most difficult aspect is the absence of a reliable radio synchrotron
spectrum. In other SNRs, such as SN 1006, Cas~A and Tycho's supernova, the
deviation of the radio photon index $\alpha$ from its test particle value
$\alpha = 0.5$ allows the determination of the injection rate of the dominant
nuclear particles into the acceleration mechanism at the forward SNR shock, and
thereby the nonlinear modification of the shock and of the acceleration
process. Fitting then the overall synchrotron spectrum, from radio to X-ray
frequencies, at the same time the effective magnetic field inside the SNR can
be determined with reasonable accuracy.

In the absence of a radio spectral index we have estimated the strength of 
the effective field by analysing the observed width of an X-ray filament 
in a smoothed radial profile which clearly indicates the position of the 
outer shock. The result is a lower limit to the effective magnetic field 
strength of about $70$~$\mu$G. Taking a value of about $130$~$\mu$G to be 
representative for the postshock value over $80$\% of the remnant's outer 
surface, we can calculate the IC \gr emission from the known X-ray 
synchrotron emission. Its negligible contribution -- at the present epoch 
-- to the observed TeV \gr flux, similar to that from nonthermal 
Bremsstrahlung, shows that SNR \rxj must be a hadronic \gr source. This is 
our first major conclusion.

The above magnetic field strength is clearly amplified with respect to its
expected value at the inner border of the hot wind bubble. Theoretically this
amplification can only come from an efficiently accelerated nuclear particle
component. And indeed, choosing as a consequence a proton injection rate
similar to those determined for the other young SNRs mentioned above, it is
possible to consistently describe the entire nonthermal broadband spectrum over
19 orders of magnitude in photon energy, including now the observed
$\pi^0$-decay \gr spectrum. This is our second important conclusion, and it is
far from trivial. On the other hand, one could argue that this remarkable
result could be achieved because the observational constraints are not as
numerous as they would be in the presence of a complete multi-wave-length
knowledge.
%
%------------------------------------------------------------------------fig.8-
\begin{figure}
\centering
\includegraphics[width=7.5cm]{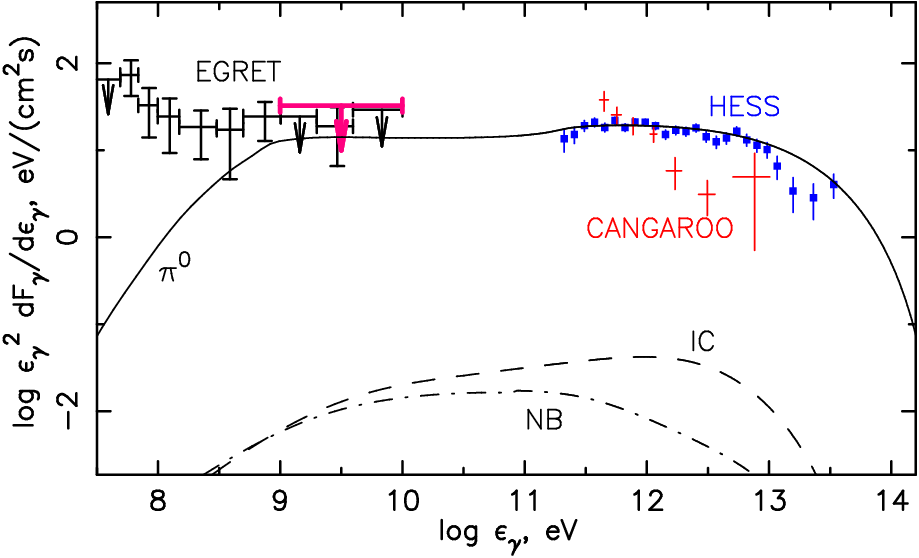}
\caption{Nonthermal Bremsstrahlung (dash-dotted line),
IC (dashed line) and $\pi^0$-decay (solid line) $\gamma$-ray
spectral energy distributions as a function of $\gamma$-ray energy for the
case, where Eq.\ref{eq15} is imposed as an additional constraint. The following
parameters have been modified compared to those on which Fig.\ref{f6} is based:
$E_{\mathrm{sn}}=2.5\times 10^{51}$~erg, $M_\mathrm{ej}=5M_{\odot}$, 
$B_0=50~\mu$G,
$\eta = 10^{-4}$, which give rise, at the current epoch, to the values:
$V_\mathrm{s}=1651$~km/s, $\sigma=4.93$, and $\sigma_\mathrm{s}=3.71$. 
$N_\mathrm{g}$ was chosen as
$N_\mathrm{g}=1.76$~cm$^{-3}$, in order to be able to visually compare the 
form of
the resulting hadronic \gr energy spectrum with the observations. The data
plotted are the same as in Fig.\ref{f6}.}
\label{f8}
\end{figure}
%------------------------------------------------------------------------------
%
It therefore remains to justify the high nuclear injection rate to positively
predict the amplitude of the \gr flux. A reasonably good radio spectrum could
achieve this. The possibility of measuring a good thermal X-ray spectrum seems
very limited from the known XMM-Newton results.

In the following we shall give a further going argument which is based on the
physical idea that the magnetic field amplification is driven by the gradient
of the CR pressure upstream of the outer shock. In fact, for all the thoroughly
studied young SNRs mentioned above, the ratio of magnetic field energy density
$B_0^2/8\pi$ in the upstream region of the shock precursor to the CR pressure
$P_\mathrm{c}$ is about the same.  Within an error of about 50 percent we have
\begin{equation} B_0^2/(8\pi P_\mathrm{c}) \approx 5\times 10^{-3}. 
\label{eq15}
\end{equation}

Requiring now this semi-empirical relation to hold also 
for SNR \rxj, we can ask ourselves, what magnetic field strength,
larger than the lower limit which we could determine from the XMM 
observations, can reproduce the observations. For this consideration we 
assume in addition that the total mechanical explosion energy 
$E_{\mathrm{sn}}$ is not larger than $2.5\times 10^{51}$~erg, i.e. that 
\rxj is not an outlier in the distribution of core collapse explosion 
energies with main sequence progenitor masses below $20 M_{\odot}$ 
\citep[compare also][]{Hamuy03}. Together with other parameters (see Fig.\ref{f8}), 
roughly equal to those given in section 2, even this large value of 
$E_{\mathrm{sn}}$ can reproduce the observations, not counting here the 
\gr observations. The associated proton injection rate is then also 
determined. Finally, the resulting theoretical hadronic \gr spectrum must 
be compared to the TeV observations. Obviously the above additional 
constraints make it more difficult to fulfil the sum total of the 
constraints. In addition, the relevant gas density at the bubble wall 
remains poorly constrained. However, the ratio $E_\mathrm{c} 
/E_{\mathrm{sn}}$ of nuclear CR energy to mechanical explosion energy is 
now limited from below.

In a limited coverage of the entire parameter space it turns out that (i) 
$B_\mathrm{d} \approx 250$~$\mu$G (ii) that the ion injection rate $\eta 
=10^{-4}$ is a factor of 3 smaller than assumed in the previous section, 
and (iii) that the total CR energy inside the remnant is $E_\mathrm{c} = 
0.22f_{\mathrm{re}}E_{\mathrm{sn}}\approx 10^{50}$~erg. It will grow in 
the spherically symmetric case to about 50\% of $E_{\mathrm{sn}}$ in 
another few thousand yrs, in general agreement with earlier studies of 
wind-SNe \citep{bv00a}. Choosing a present-day pre-shock 
density which is about 50 percent higher than assumed in section 2, one 
can even fine-tune the amplitude factor of the hadronic \gr flux to the 
observed amplitude, in order to compare with the observed spectral
form. The result is shown in Fig.\ref{f8}. It demonstrates that the new solution 
is less nonlinearly modified than before. However, the observed part of 
the spectrum is quite well reproduced. It is quite clear from Figs.\ref{f6} and \ref{f8}
that measurements of the \gr spectrum at low energies 
$\epsilon_{\gamma}\lsim 10$~GeV, which can be done with the \gr space 
telescope GLAST in the future, could provide very useful information 
about the degree of SNR shock modification.

We conclude that the present observational knowledge of SNR \rxj can be
interpreted by a source which ultimately converts about 10\% of the mechanical
explosion energy into nuclear CRs and that the observed high energy \gr
emission of SNR \rxj is of hadronic origin.

\begin{acknowledgements}
This work has been supported in part by the Russian Foundation For Basic
Research (grant 03-02-16524). EGB acknowledges the hospitality of the
Max-Planck-Institut f\"ur Kernphysik, where part of this work was carried
out. HJV is grateful to the other 
members of the H.E.S.S. collaboration for numerous
discussions about \rxj. The authors thank L. Ksenofontov for his
assistance in the preparation of this paper.
\end{acknowledgements}

\end{document}